# United States (US) Muon Collider Community White Paper for the European Strategy for Particle Physics Update

Abstract:

This document is being submitted to the 2024-2026 European Strategy for Particle Physics Update (ESPPU) process on behalf of the US Muon Collider community, with its preparation coordinated by the interim US Muon Collider Coordination Group. The US Muon Collider Community comprises a few hundred American scientists. The purpose of the document is to inform ESPPU about the US plans for Muon Collider research and development (R&D), explain how these efforts align with the broader international R&D initiatives, and present the US community vision for the future realization of this transformative project.

**Drafting Team:** M. Begel (Brookhaven National Laboratory), P. Bhat (Fermilab), N. Craig (University of California, Santa Barbara), S. Dasu (University of Wisconsin), K. DiPetrillo (University of Chicago), S. Gourlay (Fermilab), T. Holmes (University of Tennessee), S. Jindariani[1] (Fermilab), P. Meade (Stony Brook University), S. Pagan-Griso (Lawrence Berkeley National Laboratory), M. Palmer (Brookhaven National Laboratory), D. Stratakis (Fermilab)

**Endorsers:** A. Abdelhamid (University of Tennessee), D. Acosta (Rice University), P. Affleck (University of Florida), G. Agarwal (University of Notre Dame), K. Agashe (University of Maryland), P. Agrawal (University of Oxford), R. Alharthy (University of Wisconsin), B. Allmond (Kansas State University), D. Ally (University of Tennessee), G. Ambrosio (Fermilab), O. Amram (Fermilab), A. Apresyan (Fermilab), A. Apyan (Brandeis University), C. Aruta (University of Florida), C. Arzate (University of Oregon), P. Asadi (University of Oregon), S. Ashanujjaman (Karlsruhe Institute of Technology), J. Ashley (University of Tennessee), J. Backus (Princeton University), R. Bartek (Catholic University of America), A. Batz (University of Oregon), L. Bauerdick (Fermilab), C. Bell (University of Michigan), S. Belomestnykh (Fermilab), J. S. Berg (Brookhaven National Laboratory), D. Berry (Fermilab), J. Berryhill (Fermilab), S. Bhattacharya (Southern Methodist University), I. Bigaran (Fermilab and Northwestern University), V. di Benedetto (Fermilab), O. Bitter (Fermilab and Northwestern University), K. Black (University of Wisconsin), K. Bloom (University of Nebraska), S. A. Bogacz (Jefferson Lab), J. Bonilla (Northeastern University), T. Bose (University of Wisconsin), D. Bourilkov (University of Florida), G. Brooijmans (Columbia University), E. Brost (Brookhaven National Laboratory), S. Braun (University of New Mexico), D. Brown (Western Kentucky University), M. Buen-Abad (University of Maryland), J. Butler (Fermilab), C. Campagnari (University of California, Santa Barbara), M. Campana (Northeastern University), A. Canepa (Fermilab), R. Capdevilla (Fermilab), K. Capobianco-Hogan (Stony Brook University), C. Cesarotti (Massachusetts Institute of Technology), Z. Chacko (University of Maryland), P. Chang (University of Florida), S. Chang (University of Oregon), S. Chekanov (Argonne National Laboratory), Y. Chien (Georgia State University), W. H. Chiu (University of Illinois, Urbana Champaign), W. Chung (Princeton University), R. Clark (Northeastern University), S. Cousineau (Oak Ridge National Laboratory), C. Cox (University of Tennessee), M. Cremonesi (Carnegie Mellon University), C. Csaki (Cornell University), G. Cummings (Fermilab), D. Curtin (University of Toronto), S. D. Bakshi (Argonne National Laboratory), A. Datta (University of California, Los Angeles), H. de la Torre Perez (Northern Illinois University), S. Demers (Yale University), D. Denisov (Brookhaven National

---

[1] Corresponding author sergo@fnal.gov

Laboratory), R. Dermisek (Indiana University), J. Dervan (Northeastern University), A. Dhar (SLAC National Accelerator Laboratory), D. Diaz (University of California, San Diego), M. Dittrich (University of Florida), T. Du (University of Chicago), J. Duarte (University of California, San Diego), I. Dutta (Fermilab), J. Dutta (University of Oklahoma), B. Echenard (Caltech), J. Eldred (Fermilab), P. Elmer (Princeton University), G. Eremeev (Fermilab), N. Evans (Oak Ridge National Laboratory), P. Everaerts (University of Wisconsin), J. Fan (Brown University), S. Ferrante (Cornell University), S. Ferraro (Brown University), T. Figy (Wichita State University), M. Forslund (Stony Brook University), P. Fox (Fermilab), M. Franklin (Harvard University), K. Fraser (University of California, Berkeley/Lawrence Berkeley National Laboratory), A. Freitas (University of Pittsburgh), A. Gandrakota (Fermilab), A. Gaponenko (Fermilab), M. Garcia-Sciveres (Lawrence Berkeley National Laboratory), R. Garg (Stanford University), C. Geddes (Lawrence Berkeley National Laboratory), S. Gessner (SLAC National Accelerator Laboratory), E. Gianfelice (Fermilab), A. S. Giannakopoulou (Stony Brook University), S. Gleyzer (University of Alabama), A. de Gouvea (Northwestern University), H. Gray (University of California, Berkeley and Lawrence Berkeley National Laboratory), D. Green (Fermilab), L. Grossman (Lawrence Berkeley National Laboratory), Y. Grossman (Cornell), D. Guerrero (Fermilab), T. Han (University of Pittsburgh), S. Hedges (Virginia Tech), T. Heim (Lawrence Berkeley National Laboratory), M. Herndon (University Of Wisconsin), G. Herrera (Virginia Tech), C. Herwig (University of Michigan), S. Homiller (Cornell University), A. Hook (University of Maryland), A. Hoover (Oak Ridge National Laboratory), W. Hopkins (Argonne National Laboratory), M. Hostert (Harvard University), J. Howard (Kavli Institute for Theoretical Physics), J. Hoya (Argonne National Laboratory), P. Huber (Virginia Tech), R. Husain (Harvard University), B. Jayatilaka (Fermilab), L. Jeanty (University of Oregon), D. Jenkins (University of Tennessee), D. Jiang (University of Illinois, Urbana Champaign), I. Juarez-Reyes (University of Oregon), G. Kane (University of Michigan), K. Kelly (Texas A&M University), K. Kennedy (Princeton University), C. Kianian (University of Wisconsin), D. Kim (University of South Dakota), D. Kim (Argonne National Laboratory), M. Kolosova (University of Florida), K. Kong (University of Kansas), J. Konigsberg (University of Florida), S. Koren (University of Notre Dame), A. Korytov (University of Florida), A.V. Kotwal (Duke University), G. K. Krintiras (University of Kansas), K. H. M. Kwok (Fermilab), S. Lammers (Indiana University), D. Lange (Princeton University), D. Larson (Particle Beam Lasers, Inc.), M. Larson (University of Chicago), J. Lawless (University of Tennessee), C. Lee (Fermilab), L. Lee (University of Tennessee), L. Le Pottier (Lawrence Berkeley National Laboratory), I. Lewis (University of Kansas), H. Li (University of California, Santa Barbara), L. Li (Brown University), P. Li (University of Minnesota), M. Liepe (Cornell University), G. Lima (Fermilab), M. Littmann (University of Chicago), M. Liu (Purdue University), Z. Liu (University of Minnesota), A. Loeliger (Princeton University), V. Lombardo (Fermilab), S. Lomte (University of Wisconsin), M. Low (University of Pittsburgh), X. Lu (Northern Illinois University / Argonne National Laboratory), T. Luo (Lawrence Berkeley National Laboratory), N. Luongo (Argonne National Laboratory), P. Machado (Fermilab), C. Madrid (Texas Tech University), S. Malik (University of Puerto Rico), A. Mallampalli (University of Wisconsin), N. Manganelli (Northeastern University), G. Marques-Tavares (University of Utah), Z. Marshall (Lawrence Berkeley National Laboratory), V. I. Martinez Outschoorn (University of Massachusetts, Amherst), K. Matchev (University of Alabama), A. Mazzacane (Fermilab), N. McGinnis (University of Arizona), C. McLean (University at Buffalo), D. Merenich (Northern Illinois University), J. Metcalfe (Argonne National Laboratory), E. Mettner (University of Wisconsin), C. Mills (University of Illinois, Chicago), D. Minic (Virginia Tech), M. Mironova (Lawrence Berkeley National Laboratory), R. Mishra (Harvard University), C. Mitchell (Lawrence Berkeley National Laboratory), A. Mohammadi (University of Wisconsin), V. Morozov (Oak Ridge National Laboratory), H. Murayama (University of California, Berkeley and University of Tokyo), S. Nahn (Fermilab), E. Nanni (SLAC National Accelerator Laboratory), A. Narayanan (Fermilab), C. Nee (University of Wisconsin), M. Neubauer (University of Illinois, Urbana-Champaign), D. Neuffer (Fermilab), C. Ng (SLAC National Accelerator Laboratory), D. Noonan (Fermilab), Y. Nosochkov (SLAC National Accelerator Laboratory),

A. Novak (Massachusetts Institute of Technology), J. Offermann (University of Chicago), I. Ojalvo (Princeton University), Y. Oksuzian (Argonne National Laboratory), T. Oli (Argonne National Laboratory), T. Orimoto (Northeastern University), M. Othman (SLAC National Accelerator Laboratory), K. Panchal (University of Maryland), V. Papadimitriou (Fermilab), A. Parikh (Stony Brook University), K. Pedro (Fermilab), F. Pellemoine (Fermilab), C. Pena (Fermilab), G. Penn (Yale University), M. Perelstein (Cornell University), M. Peskin (SLAC National Accelerator Laboratory), A. Petrov (University of South Carolina), M. Pleier (Brookhaven National Laboratory), S. Posen (Fermilab), R. Powers (Princeton University), S. Prestemon (Lawrence Berkeley National Laboratory), M. Purohit (USC), T. Raubenheimer (SLAC National Accelerator Laboratory), J. Qiang (Lawrence Berkeley National Laboratory), L. Rainbolt (University of Chicago), C. Rasmussen (Brookhaven National Laboratory), A. Rastogi (Lawrence Berkeley National Laboratory), M. Reece (Harvard University), I. Reed (Boston University), L. Ricci (University of Maryland), C. Riggall (University of Tennessee), R. Rimmer (Jefferson Lab), B. Roberts (Lawrence Berkeley National Laboratory), B. Rosser (University of Chicago), L. Rozanov (Imperial College London), R. Ruber (Jefferson Lab), B. S Acharya (Retired), T. Sabitsch (University of Florida), M. Safdari (Fermilab), A. Safonov (Texas A&M University), D. Saltzberg (University of California, Los Angeles), D. Sathyan (Texas A&M University), H. Schellman (Oregon State University), C. Scherb (Lawrence Berkeley National Laboratory), R. Schmitz (University of California, Santa Barbara), S. Seidel (University of New Mexico), E. Sexton-Kennedy (Fermilab), V. Sharma (University of Wisconsin), V. Shiltsev (Northern Illinois University), M. Shochet (University of Chicago), I. Shoemaker (Virginia Tech), R. Simeon (University of Wisconsin), B. Simons (Fermilab/Northern Illinois University), E. Sledge (California Institute of Technology), E. Smith (Fermilab), P. Snopok (Illinois Institute of Technology), S. Snyder (Brookhaven National Laboratory), S. Spanier (University of Tennessee), G. Stancari (Fermilab), G. Stark ( University of California, Santa Cruz), N. Strobbe (University of Minnesota), S. Stucci (Brookhaven National Laboratory), J. Stupak (University of Oklahoma), R. Sundrum (University of Maryland), C. Thompson (University of Tennessee), E. Thompson (Lawrence Berkeley National Laboratory), C. Thoreson (Northeastern University), Y. Torun (Illinois Institute of Technology), C. Tosciri (University of Chicago), N. Tran (Fermilab), A. Tricoli (Brookhaven National Laboratory), C. Tully (Princeton University), A. Tuna (University of California, San Diego), I. Valenzuela Lombera (University of California, Santa Barbara), K. Van Tilburg (New York University), J. Vay (Lawrence Berkeley National Laboratory), W. Vetens (Syracuse University), C. Vuosalo (University of Wisconsin), C.E.M. Wagner (University of Chicago), B. Wang (University of Iowa), C. Wang (Fermilab), L. Wang (University of Chicago), Z. Wang (New York University), J. Watts (University of Tennessee), H. Weber (Cornell University), M. Williams (Massachusetts Institute of Technology), H. Witte (Massachusetts Institute of Technology), J. Womersley (University of Edinburgh), D. Wood (Northeastern University), Y. Wu (Nanjing Normal University), K. Xie (Michigan State University), S. Xie (Fermilab and Caltech), W. L. Xu (SLAC National Accelerator Laboratory), A. Yagil (University of California, San Diego), K. Yonehara (Fermilab), B. Yu (Cornell University), T. Yu (University of Oregon), R. L. Zamora Peinado (Virginia Tech), G. Zecchinelli (Boston University), J. Zhang (Argonne National Laboratory), Y. Zhong (City University of Hong Kong), I. Zoi (Fermilab), J. Zupan (University of Cincinnati)

# Executive Summary

The Large Hadron Collider (LHC) has had a transformative impact on high-energy physics (HEP), most notably with the discovery of the Higgs boson. This discovery without any sign for other new physics has spurred critical inquiries into the origins of electroweak symmetry and the underlying structure of the universe. In response, the 2023 US Particle Physics Project Prioritization Panel's (P5) report, developed under the auspices of the US High Energy Physics Advisory Panel, underscores the need for sustained R&D efforts to design a future 10 TeV parton center-of-momentum (pCM) energy collider, which could unlock the answers to the profound questions the LHC has begun to pose. Among the various possible collider technologies, a muon collider stands out as likely to be the most powerful tool for addressing many of the key questions raised by the LHC. Furthermore, a muon accelerator facility enables unique opportunities to probe a broad array of other scientific challenges, including dark sectors, neutrino physics, hadronic physics, charged lepton flavor violation, and more. **To fully leverage these opportunities, the US scientific community intends to build on the P5 report's recommendations by rigorously exploring the feasibility of this cutting-edge technology and its strong scientific potential to help realize our "muon shot".**

Since the latest 2021-2023 US community-wide Snowmass process, the US Muon Collider community has been working closely with international partners under the coordination of the International Muon Collider Collaboration (IMCC). Together, we have developed a detailed R&D program aimed at addressing the key technological risks associated with muon colliders. This program outlines the necessary resources and timeline for its execution. Significant technical progress has been made in Europe in the last few years, with IMCC already contributing roughly 50% of the required resources to stay on the timeline presented in this paper. The US Muon Collider community aspires to contribute at least the remaining 50%, focused on the areas the US has scientific and technical expertise and interests. A central component of the R&D roadmap is an ionization cooling demonstrator, which is essential for establishing that the luminosity goals of a muon collider are achievable. Under the foreseen timeline, the demonstrator is expected to be constructed and operational in the 2030s, and in accordance with plans envisioned by the 2023 P5 report, we hope to host this facility at Fermi National Accelerator Laboratory (Fermilab). **Assuming expanded funding and successful technological developments, we anticipate achieving technical design readiness for the collider within approximately 20 years, which could pave the way for the start of a construction project in the mid-2040s and operations commencing soon thereafter.**

It is clear that no future collider can be built without global support. If the upcoming ESPPU continues to pursue a future e+e- Higgs factory at CERN, its success will require international cooperation that includes enthusiastic participation from the US community. Similarly, a successful muon collider (wherever it is hosted) will depend on commitments from around the globe, including continued engagement from Europe and expanded engagement from the United States. We therefore support a US program that allows both of these program initiatives to co-exist, as envisioned by the P5 report.

A potential outcome of this global program is a US-hosted muon collider. This machine would enable physics exploration complementary to the Higgs Factory as well as the most efficient path to exploring nature at the highest energies. This vision is supported by the 2023 P5 report, which calls for hosting a major international collider facility in the US, one that positions the country as a leader in the global exploration of fundamental physics. **Regardless of the ultimate location, we hope that the US, Europe, and other partners from around the world can benefit from the ongoing close collaboration and mutual support, building on the lessons learned from their partnership at the LHC, HL-LHC, and neutrino programs.**

The US community is engaged and eager to ramp up work on muon colliders. The need for global muon collider R&D efforts is clear, and the community is infused with excitement, both in terms of tackling the technical challenges and the potential for a domestic siting. In August 2024, the community had its first open meeting at Fermilab, which had over 300 attendees, 260 of which were in person. Many were newcomers to the effort, looking for the best place to contribute, and over 60 people attended tutorials focused on detector and accelerator simulation software. The meeting also launched discussions of the formation of the US Muon Collider Collaboration (USMCC), which is in the process of finalizing a collaboration charter, to be adopted in 2025.

The creation of the USMCC provides an organization in the US to engage with the IMCC to jointly advance and deliver the R&D and design work required for the design of a muon collider. In the US context, its formation is a pivotal next step in advancing the P5 recommendations for a muon collider. This collaboration will define the key tasks in preparation for the mid-term review panels in the US recommended by P5, ensuring alignment with strategic goals. It will also be vital to maintain strong engagement with the international community, promoting collaboration and sharing knowledge as we move forward with muon collider research. One critical element of this effort will be the development of a US-based cooling demonstrator proposal, which will be presented to the international community for decisions on its eventual location. Additionally, a comprehensive long-term vision for Fermilab will be developed, positioning it as a central hub for muon collider R&D efforts. Such initiatives will be underpinned by a robust theoretical physics case, which will drive the scientific objectives and future discoveries that propel this groundbreaking initiative.

**The US community looks forward to continued collaboration with IMCC and strongly believes that it is essential for muon collider efforts at CERN and in Europe to receive sufficient funding and support.** An addendum agreement related to muon collider cooperation under the existing DOE-CERN cooperation agreement structure, which currently is under development, could facilitate closer coordination and establish the groundwork for additional international cooperation on this transformative project.

# Introduction

This document is being submitted to ESPPU on behalf of the US Muon Collider Community, comprising a few hundred American scientists who have voluntarily aligned themselves with the Snowmass muon collider effort by either directly contributing to R&D activities or by expressing interest through subscription to the communication channels. The interim US Muon Collider Coordination Group consists of approximately ten elected members responsible for guiding the community toward the formation of a more formal collaboration. The initial draft of the document was prepared by a drafting team and subsequently circulated among the community to solicit feedback and secure endorsement. After the initial set of comments from the community were addressed, an open meeting was held prior to submission to discuss any other remaining issues. The white paper was distributed on the US mailing lists and primarily signed by individuals from US institutions. However, about 10 signatories were from outside the US, and we chose to include them.

Particle physics has entered a new era since the discovery of the Higgs boson at the LHC. Many long held questions about the universe remain while new ones have been sparked by the LHC. We are now in search of answers to the myriad of questions surrounding why electroweak symmetry breaking (EWSB) occurs which requires studying nature at energies significantly above the electroweak scale. Furthermore studying nature at high energies opens up the possibility to observe qualitatively new phenomena predicted within the SM while simultaneously searching for new physics at the shortest distances possible. The 10 TeV scale was recently singled out by P5 as a compelling waypoint in humanity's pursuit of understanding the most basic building blocks of nature. It allows us for the first time to significantly extend into the regime of "unbroken" electroweak symmetry where possible answers to what causes EWSB could arise such as if the Higgs itself is a composite and what sets the scale of EWSB that allows for macroscopic life in our universe. Furthermore, the 10 TeV scale lets us explore the Higgs potential with precision which will shed light on the origin and fate of the universe. The 10 TeV scale also lets us test directly for the first time some of the simplest possibilities for particle dark matter based on the simplest freeze out mechanism that explains their cosmological abundance. A muon collider offers an ideal and unique path to the 10 TeV scale and beyond as the first high energy electroweak collider.

Muon colliders offer several unique advantages over hadron and electron machines. Unlike protons, muons are elementary particles where all of the beam energy becomes available to the collision. At high energies, muons enable vector boson fusion, offering an attractive opportunity to study electroweak symmetry breaking without large QCD backgrounds. Roughly 200 times heavier than the electron, muons also emit around $10^9$ times less synchrotron radiation than an electron beam of the same energy, enabling operations through compact and power efficient machines. For example, a ~10 TeV Muon Collider complex can be accommodated on the Fermilab site and could be the most efficient machine in terms of power per luminosity. Finally, a muon collider program offers unique synergies with the neutrino and precision frontiers of particle physics, via neutrino factories and fixed target experiments, and thereby bridge communities and unite them under one initiative.

This unparalleled potential has reignited significant physics interest in muon colliders over the past several years, such that the collider physics communities in Europe and in the United States have been driven to resume design studies and technological R&D. Building and operating a muon collider poses several unique technology challenges. Muons must be produced, cooled, accelerated, and brought to collide in sufficient numbers before they decay. Radiation due to the strong flux of collimated neutrinos originating from muon decays in the straight sections of the collider ring need to be managed and mitigated. Experimental detectors have to be designed to handle the diffuse high multiplicity beam-induced backgrounds from the muon decays as well. Occupancies in the detectors are expected to be similar to that of the HL-LHC; the event rates would be much smaller. Advances over the past decade have brought the necessary technology closer to realization, but targeted R&D is necessary to address engineering challenges, construct demonstrators, and make further design progress. Such an R&D program is highly attractive because it would prioritize investment into accelerator and detector development over the cost of large civil construction, maximizing potential economic, societal, and scientific benefit.

The possibility of hosting a muon collider in the US has been fully embraced by the US High Energy Physics Advisory Panel-driven P5, evident from the 2023 P5 report. In particular, P5's recommendation #4 specifically advocated for "an aggressive R&D program that, while technologically challenging, could yield revolutionary accelerator designs that chart a realistic path to a 10 TeV pCM collider." In its Executive Summary, the report explicitly recommended "targeted collider R&D to establish the feasibility of a 10 TeV pCM muon collider." Moreover, Section#2 of the report also referred to the pursuit of the path towards a muon collider by the US as our "Muon shot".

## Physics Interests

The physics case of a high energy muon collider has been rapidly developed by the worldwide theory community since the last ESPPU and Snowmass processes. Over 100 papers with "muon collider" in the title have been posted directly on the phenomenology arXiv since 2020. The driving force behind this renaissance in muon collider theory is the realistic promise of an alternative to proton colliders for probing significantly higher energies than the LHC, one that is uniquely suited to answering central questions facing the particle physics community. The discovery of a Higgs-like boson at the LHC, without so much as a hint of an explanation for the scale of Electroweak Symmetry Breaking (EWSB) or why it even occurs, requires a dedicated exploration that continues beyond future Higgs Factories (HFs). While e+e- HFs represent a technologically-ready first step after the LHC, the unique nature of the Higgs mechanism implies that certain questions can only be answered at higher energy. Furthermore, given the precision proposed at HF, whether or not deviations from the SM are found the indirect scales tested imply they can be probed directly with a 10 TeV muon collider. Therefore it is crucial to develop technologies that allow us to explore higher energies. This logic extends well beyond the Higgs boson to include many of the principal candidates for physics beyond the Standard Model (SM). For example, minimal models of weakly interacting massive particle (WIMP) dark matter require new higher-energy colliders and would be discovered by a 10 TeV muon collider, should such dark matter candidates be realized in nature. The need for a higher-energy successor to the LHC motivated the global theory community to begin surveying the possibilities for a 100 TeV pp collider long before the recent muon collider efforts. Consequently, many initial physics studies for a 10 TeV muon collider were driven by the same underlying theoretical motivations as for a 100 TeV proton collider. However, the theory case for a muon collider transcends the boundaries of the energy frontier and continues to advance further.

A muon collider offers both energy and precision, therefore a number of precision studies similar to HF and linear e+e- colliders at high energy have also been performed. The high energy of a 10 TeV muon collider generates larger cross sections for SM processes at "low energy" and enables a continuation of the Higgs precision program beyond HF. Furthermore, it offers an unprecedented window for exploring the Higgs potential with the added energy allowing for significant di-Higgs and tri-Higgs production. Moving forward, the primary focus of muon collider theory is to fully explore the unique capabilities that muon colliders provide beyond the energy frontier studies of a 100 TeV pp collider and HF. These directions follow from the ability to combine energy and precision for the first time: low backgrounds with high luminosity at the highest energies. This enables an indirect reach well beyond the scale achievable by any means for a 100 TeV pp collider in scenarios relating to extensions of the EW sector and the phenomenon of EWSB. It also allows new territory to be explored from beneath the scale of EWSB to the collider's kinematic limit in weakly coupled theories. The multipronged investigation of weakly coupled quantum possibilities takes us beyond HF as well as enabling the exploration of new phenomena within the SM. These latter directions exemplify why a muon collider is unique and more than *just* a possible route to the 10 TeV pCM scale. Muon colliders, if they can be built, are vital tools that should be exploited even if there were other collider options that could reach this energy on a similar timescale. The combination of precision with energy is just one facet of their unique physics potential.

Numerous physics cases that go beyond the standard e+e- and pp paradigms have also been investigated since P5. The most obvious case involves the synergies and potential for understanding the many mysteries surrounding neutrino physics. While advancements in neutrino physics has been long intertwined with muon colliders, the prospects for exploiting high-energy neutrino beams for new physics searches and electroweak precision measurements, the search for the mass generation mechanism, and the synergy with long baseline experiments and neutrino telescopes all must be revisited in the context of a DUNE- and HYPERK-based future. Exploration of the physics potential in other directions such as lepton and quark flavor violation is underway. Moreover, a muon collider's strength in Higgs physics gives it natural advantages in exploring the phase diagrams of EW symmetry as well as dark sectors that couple through the Higgs and other portals; the plethora of possibilities leaves much to be explored. Furthermore the development of theoretical tools and calculations to better understand and illuminate the physics case more clearly is also underway. Theory research related to a muon collider is not specific to the US community and is part of a global exponential growth seen in this direction. The US theory community will continue to support these efforts by helping to lead in new directions, and also by expanding studies in areas where the US has considerable scientific expertise such as neutrino physics and dark sectors.

# R&D Needs and potential US Contributions:

## Accelerator R&D:

Efforts in accelerator R&D can be divided into three main areas: 1) The machine design effort which will focus on the design and simulation of all the subsystems including an end-to-end performance evaluation, 2) the technology development which will focus on key areas such as targetry, magnets and radiofrequency (RF) cavity development, and finally 3) the muon cooling demonstrator program (described in a separate section below) that addresses complex integration issues.

Starting with the machine design, the US will contribute to an integrated design of all the subsystems, which needs to be developed. This includes an ionization cooling channel design that reaches the desired luminosity as well as accelerator designs that can deliver TeV scale beams with minimum losses; this is an area where the US has particularly strong interests and expertise. Once the generic machine design is established, work is also necessary to evaluate its compatibility with the US siting options.

In the baseline scheme, muons are produced as tertiary particles from the decay of pions created by a high-power proton beam impinging a high Z material target. Fermilab's existing accelerator evolution plan may offer a path to a proton driver for the muon collider. However, an R&D program must be developed to define additions to the plan that would produce protons with the desired power and time structure. The Spallation Neutron Source (SNS) accelerator at Oak Ridge National Laboratory is a close analog to the Muon Collider proton driver. It thereby offers a test-bed for demonstrating novel high-intensity technologies such as laser assisted charge exchange injection, bunch compression and space-charge mitigation techniques. Furthermore Fermilab's FAST/IOTA is a dedicated beam dynamics R&D facility that can offer detailed experimental studies on bunch compression, space-charge mitigation technology, and fast beam diagnostics for charge-dominated proton bunches.

Technology development for target survivability is a crucial area of the R&D where the US plans to contribute. In 2007, a proof-of-principle MERIT experiment at CERN validated this concept with a liquid Hg target. Recent studies indicate that more operational friendly materials such as a solid graphite, liquid metal or fluidized tungsten target are feasible options. Developing high power targets is a central focus for the Fermilab-hosted intensity frontier experiments that can be synergistic with the Muon Collider program. For example, the Mu2e experiment's geometry is a lighter analog of the muon collider target system, with targets within high field large-bore solenoids. Fermilab also hosts the RaDIATE collaboration that explores targetry challenges to support High Power Target development, such as LBNF at 2.4 MW operation. Hence collaboration with Mu2e, Fermilab's neutrino program and RaDIATE to synergistically conduct target R&D and develop the technology would strongly benefit advances for a muon collider.

In the area of RF development, the US will contribute to the development of prototype 200-800 MHz cavities for the proposed cooling channels. It is critical to study novel cavity designs or novel methods to maximize their efficiency. For example, using cryogenic copper or breakdown resilient materials such as copper alloy or aluminum. The R&D program should also investigate power sources that can deliver high-peak power (3 MW) and short pulses at hundreds of MHz. In the US, expertise exists in Normal Conducting RF (NCRF) technology in universities and the DOE national labs that could be leveraged. For example, considerable synergies exist with various efforts at SLAC, where groups have shared research interests in developing advanced NCRF cavities/structures for various applications, and in understanding the physics of RF breakdown under different conditions, including in strong magnetic fields. Muon acceleration will require high-gradient superconducting 325-1300 MHz RF cavities. The muon collider communities need further conceptual designs of the cavities as needed for accelerator lattices, followed by choosing cavities to be prototyped, creating engineering designs and preparing a prototyping program. Additional synergies exist with the DOE-HEP General Accelerator R&D program and with FCC-ee, as well as the potential for collaborations with US universities and DOE national laboratories.

The US also plans to contribute to key areas for the magnet technology development needed for a 10 TeV muon collider. Considerable advancements in the magnet design beyond the currently available technologies are needed for a 10 TeV muon collider. Three main categories of magnets have been identified as the focus of an R&D program:

1. Fixed-field superconducting solenoids for the target, decay and capture channel, 6D cooling channel and the final cooling channel.
2. Fast ramped normal conducting magnets for the acceleration rings.
3. Fixed-field superconducting accelerator magnets (dipoles, quadrupoles and combined function) for the acceleration and collider rings.

Some of these magnets can be built using $Nb_3Sn$, but others will require High Temperature Superconducting (HTS) technology development. Early R&D could leverage the US Magnet Development Program (MDP) that is generic in nature. The program has recently updated their roadmap to include solenoids and ramped up the development of magnets based on HTS. The US effort is complementary to the IMCC and European Magnet R&D programs. Muon Collider R&D can also benefit from ongoing activities at the National High Magnetic Field Laboratory, particularly with high field solenoids, and basic high field conductor and magnet R&D.

## Detector R&D:

The detector R&D can be divided into two main areas: 1) the detector design effort which will focus on design and simulation of different detector configurations and evaluation of their performances and 2) the technology development which will focus on key areas such as tracking, calorimetry, detector magnet, and data readout.

The environment around the interaction point (IP) presents unique challenges for a detector design. Dedicated shielding around the IP reduces the flux of high-energy electrons from muon decays, with still a large number of low-energy particles, mainly electrons, photons and neutrons, diffusing in the detector. The optimal shape, location, mechanical support and integration/installation will be subject to more detailed multidisciplinary studies. The detector design is heavily influenced by such a background in both its acceptance and technology choices.

High-granularity tracking and calorimetry devices with precision timing capabilities are core capabilities needed to successfully disentangle the physics objects of interest, with very different requirements on the front-end electronics compared to the LHC-era designs. It should be noted that while detector technologies exist that can reach assumed timing and feature-size specs individually, significant further development is needed to integrate, scale, and maintain low power.

Great progress has been made on a conceptual design of both the tracker and calorimeter systems, starting to explore different options and using detailed Geant4-based simulation to assess the physics performance of the conceptual detector designs. Looking ahead, significant work in exploring even more options and selecting ones that offer the most promising solutions is a key goal for the coming years. The development of a complete set of requirements in strong collaboration with accelerator design and physics projection studies is a key milestone along the way, followed by a full set of specifications to identify which of the solutions are best suited.

The US has already a very significant R&D program towards such capabilities and is eager to make progress from technology development to small-scale system prototypes to demonstrate solutions, in synergy with European efforts. In conducting this work, the community will leverage numerous synergies with the Higgs Factory and future hadron collider detectors.

Investigation of other subsystems, from muon spectrometers to PID detectors to luminosity measurement devices, to a detailed study of the solenoid are also in need of conceptual design where the US can contribute to the international effort. Globally-shared computing resources and the development of common software tools is of paramount importance to enable these studies and arrive at a full set of requirements and then specification for the detector design; current US efforts are expected to continue and ultimately grow in support of this direction.

Software and computing (S&C) are essential for detector design and performance evaluation, making the development and maintenance of the software stack a significant contribution to the overall effort. US physicists have a strong history of leading S&C initiatives in high-energy physics and have actively participated in the IMCC software task force, which is charged with planning the evolution of muon collider detector software. Looking ahead, early collaboration with the Key4hep initiative will streamline these efforts by integrating independently developed components, reducing redundant maintenance, and addressing challenges specific to the muon collider. Adopting the edm4hep data format will further enhance compatibility with modern analysis tools. Key priorities include developing 4D tracking algorithms and advancing particle flow reconstruction. As computing transitions toward GPU-based and heterogeneous systems, initiating muon collider–specific development early is crucial for addressing challenges such as beam-induced background and detector occupancy. New cross-device approaches, both for classical and machine learning algorithms, offer promising solutions for efficient data processing.

## Demonstrator Program:

The physics of ionization cooling has been demonstrated by the Muon Ionization Cooling Experiment (MICE) and the Fermilab Muon g-2 Experiment. Although the principles of ionization cooling are understood, several challenges associated with the cooling technology and its integration exist. For example, operation of NC cavities near to superconducting magnets may compromise the cryogenic performance of the magnet. Installation of absorbers, particularly using liquid hydrogen, may also be challenging in such compact assemblies. Moreover, mitigation approaches to manage the forces within and between the magnet coils need to be developed. To understand and reduce the associated risks, a facility that contains a sequence of ionization cooling cells that closely resemble a realistic ionization cooling channel is envisioned. Such a facility will allow the integrated performance of the systems to be tested as well as provide input, knowledge and experience to design a buildable cooling channel for a muon collider. In the next few years, a targeted effort should be carried out towards a conceptual design of a demonstrator for testing cooling technology as well as an effort towards exploration of candidate sites that can host such a facility.

Potential sites for the demonstrator have been identified both in the US and at CERN. In the US, Fermilab is an ideal site for hosting such a Demonstrator because of the existing accelerator infrastructure of a MW-scale proton source that currently enables a strong physics program with neutrino and muon beams, and include plans for the evolution of the complex towards a multi-MW proton beam.

## IMCC and US Engagement

Over the last few years, the US community has been building its efforts in close collaboration with the IMCC. Many US institutions have already joined the collaboration, routinely attend collaboration meetings, and contribute to the efforts of the study. The preparation of a DOE-CERN agreement concerning muon collider cooperation is now underway and would explicitly enable the participation by the DOE national labs. The IMCCcurrently has US members on its steering board, and US members have been editors on the IMCC Interim Report as well as its input to the ongoing ESPPU process. Furthermore, the leadership of the IMCC has been periodically invited to US meetings, where a strong sense of shared purpose exists across the various groups.

The existing structure of IMCC reflects the early engagement and funding in Europe. It is hosted by CERN, and closely aligned in structure with an EU-funded "MuCol" project. In order to realize the ambitious plans envisioned by P5 in the US, global support is needed. US researchers are eager to increase their participation in the IMCC, scientifically as well as organizationally. Discussions are now underway to enable more global engagement, including allowing the US and other interested countries to be better represented in the leadership of the organization.

## Timeline and Resources

While awaiting an official response from funding agencies to the relevant P5 recommendations, the US community has remained proactive. It has leveraged various funding sources, including experimental and theory-based grants, LDRD (Laboratory Directed Research and Development at the DOE national labs), university-specific funding, private sector contributions, and funding from theory institutes to support workshops and research initiatives. It is important to highlight that, at present, the majority of funding for muon collider R&D efforts comes from European sources. Therefore, it is crucial to maximize international collaboration and efforts to ensure the success of these projects.

In the course of the P5 process, the panel requested input from the US Muon Collider community about the resource allocation necessary for the muon collider accelerator and detector R&D programs. The resource projections were made with the ultimate goal of establishing the technical feasibility of a high-energy muon collider, compatible with potential US-based siting options. This input submitted to P5 is briefly summarized in this section.

The proposed muon collider R&D timeline aims to deliver a Technical Design Report (TDR) for the collider within the next 20 years, aligning with the completion of the LBNF/DUNE Phase-2 construction by the US and the international partners. This timing, coupled with the synergies and funding profile, makes the muon collider a natural candidate for a post-DUNE US-hosted flagship project in HEP. During the next two decades, the muon collider program requires moderate funding levels, which can be sustained alongside the ongoing construction of LBNF/DUNE and a Higgs Factory. The timeline presented to P5 and described in this paper differs slightly from the technically constrained IMCC timeline due to considerations regarding the availability of US resources and facilities. However, these differences are limited and result in only a short delay of a few years in the delivery of the TDR. The corresponding timeline may potentially be shortened slightly, for example depending on funding priorities after the HL-LHC accelerator and detector upgrades are completed, and therefore is well aligned with the IMCC timelines which are also being updated as part of the 2024-2026 ESPPU process.

The development of the collider and its underlying detector hardware are closely interrelated, as is the overall detector design. These elements are all intrinsically tied to the program's physics objectives. An R&D and funding strategy that recognizes and reinforces these connections will be especially effective in achieving the scientific milestones needed to demonstrate feasibility.

The program is divided into two primary phases: the **R&D Phase** and the **Demo Phase,** as shown in Figure 1. During the **R&D Phase**, which is expected to last 7-8 years, the accelerator team will focus on developing a reference design through simulation, with smaller-scale studies of key accelerator components. Concurrently, the detector team will work on designs for a 10 TeV detector, the machine-detector interface (MDI), and potential intermediate staging options. Detector component and technology R&D will also occur during this period.

By the end of the **R&D Phase**, the program aims to deliver:

1. A **Reference Design Report (RDR)** for the collider facility, which will include preliminary designs for 3 TeV and 10 TeV detectors.
2. A **TDR** with a detailed cost estimate for the demonstration facility, including a description of the physics that can be studied with the demonstrators.

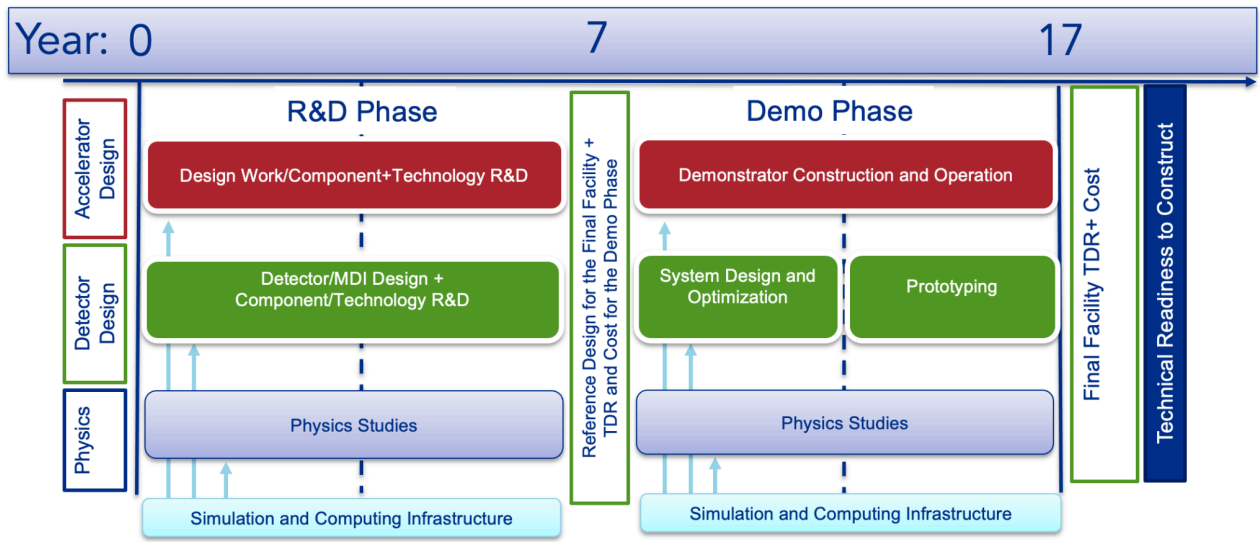

**Figure 1:** US Muon Collider R&D timeline presented to the 2023 P5. Year 0 indicates the time when significant dedicated R&D funding becomes available in the US. The R&D plan aims to achieve technical readiness for start of construction in under 20 years.

The FTE (Full-Time Equivalent) and M&S (Material & Services) needs for the **R&D Phase** have been evaluated using both bottom-up and top-down approaches. This was done under the assumption that approximately 50% of the work will be undertaken by the United States, with the remainder handled by the international partners. These estimates are currently under discussion with the funding agencies and will be updated and presented to the mid-term review panel recommended by the P5.

The **Demo Phase** will begin around Years 7-8 and last for approximately ten years. The primary goal of this phase is to construct and operate facilities needed to demonstrate the technology and to address major risks and uncertainties associated with performance of the collider. The accelerator team will focus on building and operating several demonstrators, the largest being the ionization cooling demonstration facility. The detector team will focus on system-level design, explore various options for multiple interaction points, and perform initial prototyping, including the construction of a small-scale detector to demonstrate the required background rejection capabilities with beam.

The funding needed for the **Demo Phase** will depend on the specifications and siting of the demonstrators, as well as the portion of work undertaken by the United States. A more detailed budget profile for the accelerator demonstrators will be developed at the end of the **R&D Phase**, but we estimate that the peak M&S funding for this phase will need to grow by approximately a factor of 10 compared to the **R&D Phase**, to enable construction of the ionization cooling demonstrator. The detector R&D budget is also expected to grow by a factor of 2-3 compared to the **R&D Phase**, based on experience with ATLAS and CMS early-stage R&D, construction, and upgrades. These estimates are similar to those from IMCC, and US and European proponents are maintaining communication around these estimates to ensure compatibility.

The deliverable at the end of the **Demo Phase** will be a TDR for both the collider facility and the associated detectors, complete with a detailed cost estimate. We expect that during the later stages of the **Demo Phase,** the process of final feasibility study and international negotiations will be launched, ahead of the decision to proceed with the construction.

Realization of the plan above relies on sustained and adequate funding in both the US and Europe, and a close collaboration between the two regions, as outlined in the next section. Potential contributions from other regions of the world would further strengthen the program.

# Path Forward for the US

The most effective path forward for developing the conceptual design of the Muon Collider and its associated detectors in the US will involve working closely with the international community. The USMCC is currently being formed, which is an organization of individuals interested in bolstering the science case, designing, and promoting the construction of a multi-TeV muon collider. The organization will bring accelerator, experimental, and theoretical physicists together, integrating new collaborators into the muon collider effort, and organizing work towards maximal impact. The USMCC will lead the work related to siting a muon collider in the United States, while also working with the IMCC in realizing a muon collider, wherever it is built in the world.

The Muon Smasher's Guide and other studies from the theoretical particle physics community have had a significant impact on the growing interest in 10-TeV scale physics. There is a strong case for further developing this physics program with greater realism and rigor in event generation and simulation, as well as expanding to include additional areas of physics. Achieving this will require a close collaboration between phenomenologists and experimentalists. We expect that these efforts will yield valuable contributions, helping to establish prospects for measuring standard model Higgs parameters and exploring new physics sectors.

US groups, in collaboration with IMCC, have already started detector studies for a 10-TeV machine. Key aspects such as detector technologies, size, overall geometry, and segmentation are being optimized. These ongoing studies provide a strong foundation, but further work is needed to gain a deeper understanding of the critical factors and conduct a comparative evaluation to develop a viable detector proposal. A particularly important focus is mitigating beam-induced background (BIB). A close collaboration amongst accelerator and simulation experts can yield a more streamlined framework.

The US-based MAP program made significant contributions to the Muon Collider concept. Recently, the IMCC has taken the lead in Muon Collider R&D, following the US divestment from the MAP program. However, much of the expertise at DOE national labs remains intact and could be quickly revitalized to form a strong R&D team. Rebuilding this team would enable the US to participate in the recently funded CERN-based demonstrator efforts. Furthermore, reestablishing the accelerator R&D team will position the US Community to prepare proposals for the US hosted demonstrator once the DOE determines the budget scale.

The US Muon Collaboration, which will include participants from MAP and an expanding group of scientists interested in the Muon Collider, presents an excellent opportunity to design a collider for potential US siting. Fermilab stands out as an ideal location for a Muon Collider with a center-of-mass energy target at the desired 10-TeV scale. The synergy with Fermilab's existing and planned accelerator complex, along with its neutrino and charged lepton flavor violation physics programs, further strengthens the case for this investment. Eventually the IMCC and other international committees intend to form a global consensus on the actual siting and design selection.